\documentstyle[preprint,aps]{revtex}

\begin{document}
\tolerance=10000
\hfuzz=5pt
\preprint{HD-TVP-99/02}
\title{The phase diagram and bulk thermodynamical quantities
  in the NJL model at finite temperature and density}
\author{T. M. Schwarz, S. P. Klevansky and G. Papp\thanks{%
on the leave from HAS Research Group for Theoretical Physics, 
E\"{o}tv\"{o}s University, Budapest, Hungary}}
\address{Institut f\"{u}r Theoretische Physik, Universit\"{a}t Heidelberg,\\
Philosophenweg 19, D-69120, Heidelberg, Germany.}
\date{\today}
\maketitle
\begin{abstract}
We reexamine the recent instanton motivated studies of Alford, Rajagopal
and Wilczek, and Berges and Rajagopal in the framework of the standard
SU(2)
Nambu-Jona-Lasinio model. The chiral phase diagram is calculated in the
temperature--density plane, and the pressure is evaluated as the function
of the quark density. Obtaining simple approximate relations describing the
$T$-$\mu$ and $T$-$p_F$ phase transition lines we find that  the results of
the instanton based model and that of the NJL model are identical.
The diquark transition line is also given.
\end{abstract}
\vfill

\section{Introduction.}

Recent studies by several authors  using an
effective 4-fermionic interaction 
 between quarks~\cite{wilczek,berges,shur} or direct instanton 
approach~\cite{rapp}
have rekindled interest in 
the two flavor QCD phase transitions. In particular, Alford, Rajagopal and
Wilczek~\cite{wilczek} have studied the pressure density and gap parameter
using a fermionic Lagrangian with an instanton motivated four-point
interaction. At zero temperature these authors found negative 
pressure for a certain range of the Fermi momentum, $p_{\rm F}$ and
showed the solutions of the gap equation as a function of $p_{\rm F}$.
Berges and Rajagopal~\cite{berges} extended this work and have calculated 
the phase diagram of a
strongly interacting matter as a function of temperature and baryon number
density in the same model.

The question that we raise and examine in this paper is whether or not these
results are fundamentally  different from those obtained via the
standard well-known Nambu-Jona-Lasinio (NJL) 
model~\cite{NJL,sandi,vogl,hats,chris,alk}, at least with regard to
chiral symmetry breaking. 
Quantities such as the gap parameter, pressure density and other 
thermodynamical
quantities have been extensively studied in this model over the last
decade~\cite{sandi,hats,hue,zhu}, and even to a level of sophistication
that goes beyond the standard mean field treatments~\cite{hue,zhu}. However,
the results have usually been presented as a function of the chemical
potential, and not as of the Fermi momentum $p_{\rm F}$ or density, as
the authors of~\cite{wilczek,berges} have done, and hence the connection 
between their results and those of the NJL model are not obvious.
 Thus, in order to make a systematic comparison, we
have to reevaluate the gap, pressure and phase diagram in these variables.
In section II we derive a simple analytical approximate expression for
the phase boundary that is independent of the parameters of the NJL model.
In section III, we examine the pressure as a function of the density  and calculate
the
complete chiral phase diagram numerically from Maxwell constructions.   We
compare our results with that of Ref.\cite{berges}.    In Section IV, we
write down the form of the gap equation for a superconducting
 diquark transition, and
thus the line of points in the $T$-$\mu$ plane.   This is evaluated
numerically.  We summarize and conclude in Section V.
 

\section{Phase boundary curve - an analytic expression.}

We commence by deriving an approximate analytic expression for the phase
boundary curve. Our starting point is the gap equation for the dynamically
generated $up$ and $down$ quark mass $m$ that is derived from the SU(2) 
chirally symmetric Lagrangian,
\begin{eqnarray}
{\cal L}_{\rm NJL}={\bar\psi}i{\not\!\;}\partial\psi+
G[(\bar{\psi}\psi)^2+(\bar{\psi}{\rm i}\gamma_5\,\vec{\tau}\psi)^2]
\end{eqnarray}
with $G$ a dimensionful coupling and $\psi$ quark spinors for $u$ and $d$ 
quarks. At non-zero temperature and chemical potential, the mean-field
self-energy or gap equation reads \cite{sandi}
\begin{eqnarray}
\label{sigi}
\Sigma^\ast&=&m=
4GN_cN_fm\int\!\frac{d^3p}{(2\pi)^3}\frac{1}{E_p}\left[
1-f^+(\vec p,\mu)-f^-(\vec p,\mu)\right]
\end{eqnarray}
with the Fermi distribution functions
$
f_p^\pm\equiv{1}/{[e^{\beta (E_p\pm\mu)}+1]}=f^\pm(\vec p,\mu)
.
$
The gap equation is easily seen to minimize the thermodynamical potential
\cite{sandi,asakawa},
\begin{equation}
\Omega(m) = \frac {m^2}{4G} - \gamma\int\frac{d^3p}{(2\pi)^3} E_p -
\gamma T\int \frac{d^3p}{(2\pi)^3} \log
[1+e^{-\beta(E_p+\mu)}][1+e^{-\beta(E_p-\mu)}],
\label{omega}
\end{equation}
where $E_p^2=p^2+m^2$, $\beta=1/T$ and $\gamma=2N_cN_f$ is the degeneracy
factor. The condensate is related to $m$ via 
$m=-2G\langle\bar\psi\psi\rangle$.
The three momentum integrals are understood to be regulated by a cutoff
$\Lambda$, and a standard set of parameters, 
$\Lambda=0.65\,$GeV 
and $G=5.01\,{\rm GeV}^{-2}$ are used to fix the values of \mbox{$f_\pi
  =93\,$MeV} and the condensate density per flavor, $\langle\bar uu
\rangle = \langle\bar dd
\rangle =(-250\,$MeV$)^3$.

Let us examine first the $T=0$ limit of Eq.(\ref{sigi}). The gap equation
for the non-trivial solution is
\begin{eqnarray} 
\frac{\pi^2}{2GN_cN_f}=\int_{p_{\rm F}}^\Lambda\!{\rm
  d}p\frac{p^2}{\sqrt{p^2+m^2}}
\end{eqnarray}
in terms of the Fermi momentum $p_{\rm F}$. $p_{\rm F}$ is a decreasing 
function
of the constituent mass taking its maximum value, $p_c$ at the chiral phase
transition, $m\to0$\footnote{Here we assume a second order transition
where the smooth vanishing of the order parameter signals the
transition. However, for small temperatures the phase transition is 
first order with a jump in the order parameter. Nevertheless, the
approximation made here qualitatively gives the proper answer (cf. Fig.~2.).}
\begin{eqnarray}
\label{pc1}
p_{\rm c}=\Lambda\sqrt{1-\frac{\pi^2}{G\Lambda^2N_cN_f}}
.\end{eqnarray}
Using the numerical values of $G$ and
$\Lambda$ given above and fixing $N_c=3$, $N_f=2$ leads to the numerical value
$p_{\rm c}=0.307\,{\rm GeV}$.
We use this relation to eliminate $\Lambda^2$ in favor of $p_{\rm c}^2$.
Now, in order to determine the phase transition line, we start with 
the gap equation~(\ref{sigi}) and  divide $m\ne 0$ out.
The critical values of the temperature and chemical potential that lie on
the phase transition boundary are determined through the vanishing chiral
condensate and hence $m\to 0$. Performing the usual substitution 
$u=\beta(p\pm \mu)$ as appropriate, and  
evaluating the integral on the right hand side, one finds
\begin{equation}
\label{phasemu}
\mu^2=p_{\rm c}^2-\frac{\pi^2T^2}{3} \,.
\end{equation}
This equation defines the chiral phase transition curve in the 
$T$-$\mu$ plane, and in this form
contains no explicit model dependence on $G$ and $\Lambda$. We would however, 
prefer to display the phase transition line
as a function of temperature and quark density. To do so, we express the density
through the Fermi momentum, $p_{\rm F}$ at zero temperature,
\begin{equation}
n=\frac{2N_cN_f}{6\pi^2}p_{\rm F}^3 \,.
\end{equation}
At finite temperature,  
the quark density $n$ is given as 
\begin{equation}
n=2N_cN_f\int\frac{{\rm d}^3p}{(2\pi)^3}\left[f_p^--f_p^+\right] \,.
\label{density}
\end{equation}
The cube root of the density defines $n_3$ as $n^{1/3} = cn_3$, where
$c=2/\pi^2$, and $n_3$ is evaluated to be
\begin{equation}
n_3=\left[3\sinh\left(\frac{\mu}{T}\right)\cdot\int\!{\rm d}p\,
\frac{p^2}{\cosh\left({\mu}/{T}\right)
+\cosh\left({E_p}/{T}\right)}\right]^{1/3} \,.
\end{equation}
Using this transformation, we can display all quantities as
functions of $n_3$  (or density) and not of 
the chemical potential, $\mu$ any more. One can take a set of $T$ and $\mu$ 
and calculate for
example $m(T,\mu)$ and $n_3(T,\mu)$. The result obtained  can be seen in
Fig.~1 where for fixed $T$, $T=0$ the mass $m$ is plotted as a function of
$p_{\rm F}$
 (solid line) and of $\mu$ (dashed line) for comparison. We note that
the
 behavior found is
qualitatively the same as that in~\cite{wilczek}.
At the chiral phase transition point, $m\to 0$ and $n_3\to n_{3,c}$ 
with
\begin{equation}
n_{3,c}^3=\mu(\mu^2+\pi^2T^2)
.\end{equation}
This expression can be used to eliminate $\mu$ from Eq.~(\ref{phasemu})
and the chiral phase transition curve in the $T$-$n_3$ plane 
 is determined from the simple analytical expression,
\begin{eqnarray}
\frac{4\pi^6}{27}T^6-\pi^2p_c^4T^2+\left(p_{\rm F,c}^6-p_c^6\right)=0
.
\label{e:anal}
\end{eqnarray}

The solution of this third order equation in $T^2$, defining the
critical temperature, $T_c$ at which the phase transition occurs for a
given ratio $\rho\!=\!n_{\rm 3,c}/p_c$ can be written as
\begin{eqnarray}
T_c(\rho)&=&\frac{\sqrt{3}p_c}{\pi}\sqrt{\cos{\left(
\frac{1}{3}\kappa\rho\right)}} \, \le \, \frac{\sqrt{3}}{\pi} p_c
\end{eqnarray}
with
\begin{equation}
\,\kappa=\left\{\begin{array}{ll}
{\rm tan}^{-1}\sqrt{\left[\rho^6-1\right]
^{-2}-1}  & \rho\leq 1\\
\pi+{\rm tan}^{-1}\sqrt{\left[\rho^2-1\right]
^{-2}-1}  \hspace*{10mm} & 1\leq\rho\leq 2^{1/6}\end{array}\right.
.\end{equation}
The trivial solution to the gap equation is the only solution if the 
temperature exceeds
 the maximum value of the critical temperature on the
transition curve,
\begin{equation}
T_m=\frac{\sqrt{3}p_c}{\pi}\simeq 0.169\,{\rm GeV}
\label{tm}
\end{equation}
or if $n_{3,c}/p_c\geq 2^{1/6}$.
One can now recast the solution for $n_{3,c}$ in terms of $T_m$
yielding a parameter-free normalized relation,
\begin{eqnarray}
\left(\frac{n_{3,c}}{p_c}\right)^3=\left[2
\left(\left(\frac{T}{T_m}\right)^2
+1\right)\right]\sqrt{1-\left(\frac{T}{T_m}\right)^2}
.\end{eqnarray}
We illustrate this relation in a more usual way in Fig.~2, plotting the
critical value of the temperature as a function of the
quark density divided by normal nuclear matter density 
$n_0=0.17\,{\rm fm}^{-3}$.

\section{Phase transition curve via Maxwell construction.}

We return now to the thermodynamical quantities.
From Eq.(\ref{omega}) for $\Omega(m)$, it follows that the pressure density is
given as
\begin{eqnarray}
p&=&-\Omega
=\frac{\gamma}{\beta}\!\int\!\!\frac{{\rm d}^3p}{(2\pi)^3}\ln\left[
\left(1+e^{-\beta(E_p+\mu)}\right)\!\!\left(1+e^{-\beta(E_p-\mu)}\right)\right]
+\gamma\int\!\frac{{\rm d}^3p}{(2\pi)^3}E_p-\frac{m^2}{4G}\,.\nonumber
\end{eqnarray}
\vspace*{-.55in}
\begin{equation}
\end{equation}
The energy density is found to be
\begin{eqnarray}
\label{frennichtren}
\epsilon=\gamma\!\int\!\!\frac{{\rm d}^3p}{(2\pi)^3}E_p\left[f_p^-+f_p^+\right]
-\gamma\int\!\frac{{\rm d}^3p}{(2\pi)^3}E_p+\frac{m^2}{4G}
\,.\end{eqnarray}
 In the limit $T\!\to\!0,\mu\!\to\!0$, one has
\begin{eqnarray}
\label{epsvac}
\epsilon_{\rm vac}=\epsilon_{T\!\to 0,\mu\!\to 0}=\frac{{m^\ast}^2}{4G}
-\frac{N_cN_f}{\pi^2}\int_0^\Lambda{\rm d}p\,p^2\sqrt{p^2+{m^\ast}^2}
,\end{eqnarray}
where $m^\ast=m(T\!=\!0,\mu\!=\!0),$
while an  analogous calculation for the pressure density $p$ 
at $T\rightarrow 0$, $\mu\rightarrow0$ yields
\begin{equation}
p_{\rm vac}=-\epsilon_{\rm vac} \,.
\end{equation}
$p_{\rm vac}$ and $\epsilon_{\rm vac}$ are independent of temperature and
chemical potential and their value is (up to the sign) the same. For our
choice of parameters, $\epsilon_{\rm vac}=-(407{\rm MeV})^4$.
Measuring the pressure and energy densities relative to their vacuum values,
we have
\begin{eqnarray}
\label{epsfinal}
\epsilon_{\rm phys}&=&\epsilon-\epsilon_{\rm vac} \,,\\
p_{\rm phys}&=&
\label{pfinal}
p+\epsilon_{\rm vac}
\,.\end{eqnarray}
In Fig.~3, we plot the pressure density as a function of $n_3$ for a range
of temperatures.   Note that at $T=0$, $p$ becomes {\it negative} and
displays a cusp like structure.   This is bought about by the fact that the
two solutions for the gap equation enter into the pressure density on the
different arms of the pressure:  the rising curve is bought about by the $m=0$
solution, while the solution that goes down has a value of $m\ne0$.
Note that this situation is similar to that observed in Ref.~\cite{wilczek}.

The phase transition curve can now be calculated using a more standard
but numerical treatment and compared to the approximate (second order) 
curve shown as the dashed line in Fig.~2.    The difference between these
two curves lies in the fact that the analytic one is calculated with
$\Lambda\to \infty$, while the numerical curve has $\Lambda$ finite.
It is worth noting that the
truncation of the NJL model to zero modes~\cite{jnpz} shows a similar
behavior to the one observed in the full model, however, lacking the
unstable backbending of the phase curve at high temperatures.
Plotting the pressure rather as a function of volume allows one to perform
Maxwell constructions and obtain the full information of the phase diagram,
including the metastable region.   The results of this calculation are
show in Fig.~4 as a function of $n^{1/3}$ as the solid curves.   Also shown
(dashed curves) are the calculations of Ref.~\cite{berges}.   
We note that the phase transition curve for the chiral 
transition are qualitatively identical.    The main difference between these
curves lies in the observation that the mixed phase given by
Ref.~\cite{berges} starts already at $n=0$, similarly to the truncated NJL
calculation~\cite{jnpz}.   The equivalence of these  models is perhaps 
not simply apparent when one examines the Lagrangians.   However, the
thermodynamical potential of Ref.~\cite{berges} is precisely that of
Eq.~(\ref{omega}) in the absence of the diquark condensate.   Thus the
differences observed in Fig.~4 can only be attributed to the use of slightly
different parameters, plus the different method of implementing
regularization - in the NJL model here, a `hard' cutoff $\Lambda$ is
employed, and in the approximate expression for the phase curve, is also
taken as $\Lambda\rightarrow\infty$,
 while the authors of \cite{wilczek,berges} use a soft form 
factor $F(p) = \Lambda^2/(p^2+\Lambda^2)$ to regulate their momentum
integrals.    The physics however cannot and does not depend on this and 
 the qualitative results remain unchanged.

Limitations and difficulties of this model in describing thermodynamical
quantities are well-known~\cite{zhu,bub}.   We illustrate one problem in 
showing the energy per quark plotted at $T=0$ as a function of the
density in Fig.~5.
 This quantity
does not posses a minimum at normal nuclear matter density
as expected, in apposition to recent  linear sigma model 
calculations~\cite{pirner}

\section{Diquark condensate transition line}

The thermodynamic potential for three colors\footnote{for two colors a
massless pionic diquark may form.} $\Omega(m)$ given in Eq.(\ref{omega})
corresponds precisely to the $\Delta=0$ limit of the function
$\Omega(m,\Delta)$ \cite{berges}
\begin{eqnarray}
\Omega(m,\Delta) &=& \frac {m^2}{4G} + \frac {\Delta^2}{4G_1} -2N_f\int\!
\frac{d^3p}{(2\pi)^3}\Big\{(N_c-2)[E_p+T\ln(1+e^{-\beta(E_p
-\mu)})(1+e^{-\beta
(E_p+\mu)})] \nonumber \\
&+& \sqrt{\xi_+^2+\Delta^2} + s\sqrt{\xi_-^2+\Delta^2}
+ 2T\ln(1+e^{-\beta\sqrt {\xi_+^2+\Delta^2}})(1+ e^{-\beta s\sqrt
{\xi_-^2+\Delta^2}}) \Big\},
\label{e:omextend}
\end{eqnarray}
where $\xi_\pm = E_p \pm \mu$, and $\Delta$ is the color
superconducting
condensate in the diquark channel.   $s$ is a sign function, and $s=\pm1$
for
$E_p>\mu$ or $E_p<\mu$. Here the expression from \cite{berges} has been modified to
remove the form factor --  instead a 3-D cutoff $|\vec p|<\Lambda$
 as is usual in the NJL model and which was used in the previous section
is to be understood.
$G_1$ is a new coupling strength mediating a four-fermion interaction that is
attractive in the diquark channel.    The form for $\Omega(m)=\Omega(m,0)$ from
Eq.(\ref{omega}) thus 
indicates that
the models are the same, and the same gap equation for the chiral transition
is retained.    For the superconducting sector, the gap obtained by
differentiating $\Omega(0,\Delta)$ with respect to $\Delta$ is
\begin{equation}
\frac 1{2G_1} = 2N_F\int\frac
{d^3p}{(2\pi)^3}\Big\{\frac1{\sqrt{\xi_+^2+\Delta^2}}
\tanh \frac{\beta}2\sqrt{\xi_+^2+\Delta^2} + \frac1{\sqrt{\xi_-^2+\Delta^2}}
\tanh \frac{\beta}2\sqrt{\xi_-^2 + \Delta^2}\Big\}.
\label{e:supergap}
\end{equation}
This expression is a relativistic generalization of the superconducting gap
equation for electron pairs \cite{fetter}
in which the quasiparticle energies $\sqrt{\xi_-^2 + \Delta^2}$ and 
$\sqrt{\xi_+^2 + \Delta^2}$ relative to the Fermi surface are introduced.
Note that at $T=0$, one recovers the result of \cite{wilczek} for $\mu\ne0$,
assuming that the form factor of these authors is set to one.   The phase
transition line in the $T$-$\mu$ or $T$-$n_3$ planes
 can now be obtained quite simply.   Assuming that the superconducting phase
transition can only occur in the region where chiral symmetry is restored,
we may set $E=p$. Furthermore we assume the transition to be 
second order, driven by the condition $\Delta\to0$. 
Thus the $T$-$\mu$ critical curve satisfies
\begin{eqnarray}
\frac {\pi^2}{2N_fG_1} &=& \int_0^\Lambda dp p^2 \frac 1{p+\mu} 
\tanh\frac 12
\beta(p+\mu) + \int_\mu^\Lambda dp p^2\frac 1{p-\mu} \tanh\frac 12\beta(p-\mu)
\nonumber\\
&+&\int_0^\mu dp p^2\frac 1{\mu-p} \tanh\frac 12\beta(\mu-p), 
\label{e:supertrans}
\end{eqnarray}
which, with obvious changes of variables reduces to
\begin{equation}
\frac{\pi^2}{2N_fG_1} = \int_0^{\Lambda+\mu} d\xi (\xi - 2\mu
+\frac{\mu^2}\xi)\tanh\frac12\beta\xi + \int _0^{\Lambda-\mu} d\xi (\xi +
2\mu + \frac{\mu^2}{\xi})\tanh\frac12\beta\xi.
\label{e:tosolve}
\end{equation}
Unlike the case for electron pairing, in which $\mu\gg \omega_D$, with
$\omega_D$ the Debye frequency, we cannot regard the logarithmic term as
being leading, and it is not possible to obtain a simple analytic expression
for the right hand side of Eq.(\ref{e:tosolve}).   We thus solve
Eq.(\ref{e:tosolve}) numerically for the critical line.    Clearly this 
depends on the choice of the strength $G_1$ and is a sensitive function
thereof.  Arbitrarily demanding that $T_c=40$MeV at $\mu=0.4$GeV close
to the values of Ref.~\cite{berges} sets
$G_1=3.10861$ GeV$^{-2}$, or $G_1\Lambda^2 = 1.31$.   The resulting curve is
indicated by the dotted line in Fig.~4.   As can be seen, the qualitative
behavior of the model of Ref.\cite{berges} is confirmed. We found
a somewhat lower gap, $\Delta$ being $\sim 35 MeV$ at the chiral
transition point and zero temperature
increasing up to $\sim 95 MeV$ at and $\mu=0.53 GeV$ and zero temperature.
The general behavior of the gap parameter as a function of the chemical
potential and temperature is given in Fig.~6. 

Finally, we comment that although the diquark phase transition line was
investigated here under the expectation that chiral symmetry is restored,
this is not necessarily the case:  in principle, the diquark phase
transition line can extend into the region in which chiral symmetry is
broken, i.e. where $\langle\bar\psi\psi\rangle \ne 0$ or $m\ne 0$.    In
practice, this turns out to be a function of the parameters chosen.   If
the diquark phase transition line enters into the region of chiral symmetry
breaking at a temperature larger than the tricritical temperature, the
dependence of $m(T,\mu)$ that enters into Eq.(23) is continuous, and a
solution to this equation can be found.   For our choice of $G_1$, however,
the diquark transition line  would enter into the region where a 
first order phase transition takes place.   Thus $m(T,\mu)$ is discontinous,
and no physically accessible solution to this equation can be found.
The situation is illustrated in Fig.~7, in which Eq.(23) is
inverted and  solved for
$G_1$ and the functional dependence, as well as the numerical parameter
value $3.1$ GeV$^{-2}$  are plotted as a function of temperature for
several values of  $\mu$.
    One sees from
this figure that the line $G_1=3.1$GeV$^{-2}$ cannot intersect the curve
$\mu=0.3$GeV for example, 
which lies in the broken phase, and therefore there is no
physically attainable solution.   Note however
 that by adjusting the value of
the constant $G_1$ to a somewhat smaller value could admit a solution 
within this region. 

\section{Conclusions}

In analyzing the chiral phase transition in the NJL model at finite
temperature and density, we find the same behavior for the chiral and
diquark phases as that reported in 
\cite{wilczek} and \cite{berges}, which  use an instanton
motivated interaction that is also four point in nature.  
That this must occur can be seen directly from the 
explicit form of the thermodynamical potential that is well-known in our
case \cite{sandi,asakawa}, and which is obtained from \cite{berges} on 
setting the form factor to one and introducing a 3-D cutoff $\Lambda$.   In
addition, we are easily able to give an approximate analytic form for the
chiral phase curve in the $T$-$\mu$ and $T$-$n_3$ planes that is independent of
the model parameters.   We have examined the extended form of the 
thermodynamic potential that makes provision for a diquark condensate and
obtained the appropriate critical line in the NJL model.   Our qualitative
results conform with those of \cite{berges}.   In addition, we find evidence
for the appearance of a diquark condensate also within the region where
chiral symmetry is not restored, but this is strongly parameter dependent.

\section*{Acknowledgments}

One of us,
G.P., thanks Michael Buballa and Maciej A. Nowak for the discussions and
comments.
This work has been supported  by the German 
Ministry for Education and Research (BMBF) under contract number 06 HD 856,
and by the grant OTKA-F019689.

\newpage
\centerline{\bf Figure Captions}
{\bf $\\$Figure 1:} 
The gap parameter $m$, shown both as a function of $\mu$  (solid curve)
and $p_F$ (dotted line) at $T=0$.\\\\
{\bf Figure 2:} 
The phase diagram calculated from the approximate analytically form
Eq.(\ref{e:anal}) (dashed line) and as
determined numerically (solid lines) as a function of the quark density $n$,
scaled by normal nuclear matter density $n_0=0.17$ fm$^{-3}$.\\\\
{\bf Figure 3:} 
Pressure shown as a function of $n_3=(2/\pi^2)n$ (left
graph) and as a function of the volume, normalized by a characteristic
volume. \\\\
{\bf Figure 4:}
Direct comparison of the NJL phase diagram (solid lines) shown as a function
of $n^{1/3}$, with that of 
\cite{berges} (dashed lines). The lines to the right of $n^{1/3}\sim 0.19$
are the  superconducting
transition lines.\\\\
{\bf Figure 5:} The energy per quark, shown
 as a function of $n/n_0$. The 
minimum occurs at approximately
 $n\simeq 5n_0$ and not  at $n=3 n_0$\\\\
{\bf Figure 6:}  The diquark gap parameter $\Delta$, shown as a function
of $\mu$ and $T$.   The contour values are given in MeV.\\\\
{\bf Figure 7:}  The functional dependence of $G_1$ as calculated from
Eq.(23) is plotted as a function of the temperature for different values
of the chemical potential.   The curves, taken from the uppermost one,
correspond to the values of $\mu=0;0.1;0.2;0.25;0.3;0.35$ and $0.5$ GeV.

\newpage

\begin{thebibliography}{99}
\bibitem[1]{wilczek} M. Alford, K. Rajagopal und F. Wilczek,  
Phys. Lett. {\bf B422} (1998) 247
\bibitem[2]{berges} J. Berges, K. Rajagopal,
Nucl. Phys. {\bf B538} (1999) 215
\bibitem[3]{shur} E. Shuryak,
hep-ph/9903297
\bibitem[3]{rapp} R. Rapp, T. Sch\"afer, E. Shuryak and M. Velkovsky,
Phys. Rev. Lett. {\bf 81}, 53 (1998); hep-ph/9904353.
\bibitem[4]{NJL} Y. Nambu and G. Jona-Lasinio, 
Phys. Rev. {\bf 122}, 345 (1961); {\bf 124}, 246 (1961).
\bibitem[5]{sandi} S.P. Klevansky, 
Rev. Mod. Phys. {\bf 64}, 649 (1992).
\bibitem[6]{vogl} U. Vogl and W. Weise, 
Prog. Part. Nucl. Phys. {\bf 27}, 91 (1991).
\bibitem[7]{hats} T. Hatsuda and T. Kunihiro,
Phys. Rep. {\bf 247}, 241 (1994).
\bibitem[8]{chris} C.V. Christov et al., 
Prog. Part. Nucl. Phys. {\bf 37}, 91 (1996).
\bibitem[9]{alk} R. Alkofer, H. Reinhardt and H. Weigel, 
Phys. Rep. {\bf 265}, 139 (1996).
\bibitem[10]{hue} J. H{\"u}fner, S.P. Klevansky and P. Zhuang,
Ann. Phys. {\bf 234}, 225 (1994).
\bibitem[11]{zhu} P. Zhuang, J. H{\"u}fner and S.P. Klevansky,
Nucl. Phys. {\bf A576}, 525 (1994).
\bibitem[12]{jnpz} R.A. Janik, M.A. Nowak, G. Papp and I. Zahed,
Nucl. Phys. {\bf A642}, 191 (1998).
\bibitem[13]{bub} M. Buballa,
Nucl. Phys. {\bf A611}, 393 (1996).
\bibitem[14]{asakawa} M. Asakawa and K. Yazaki, Nucl. Phys. {\bf A 504},
668
(1989).
\bibitem[15]{pirner} J. Meyer and H.-J. Pirner, private communication
\bibitem[16]{fetter} A.L. Fetter and J.D. Walecka, {\it Quantum Theory
of Many-Particle Systems},\ (McGraw-Hill, New York, 1971)
\end{thebibliography}
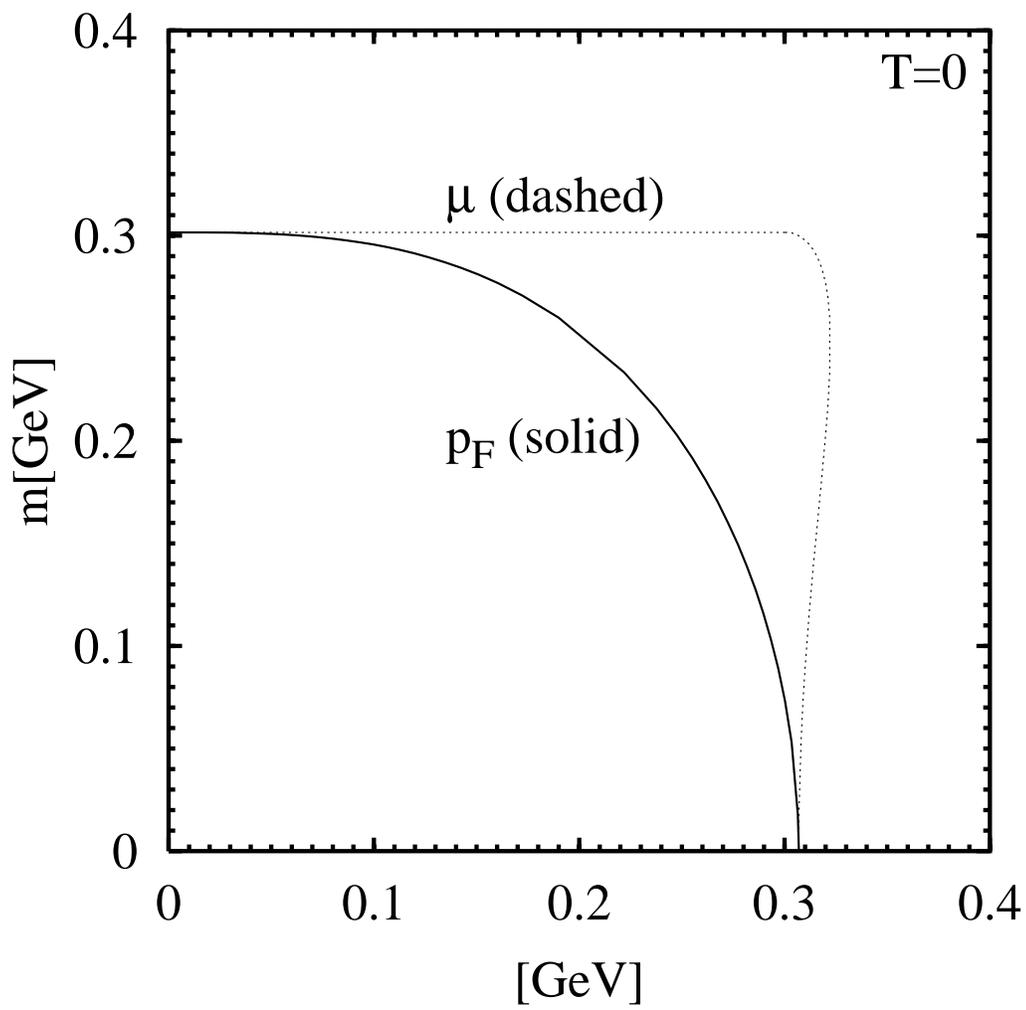
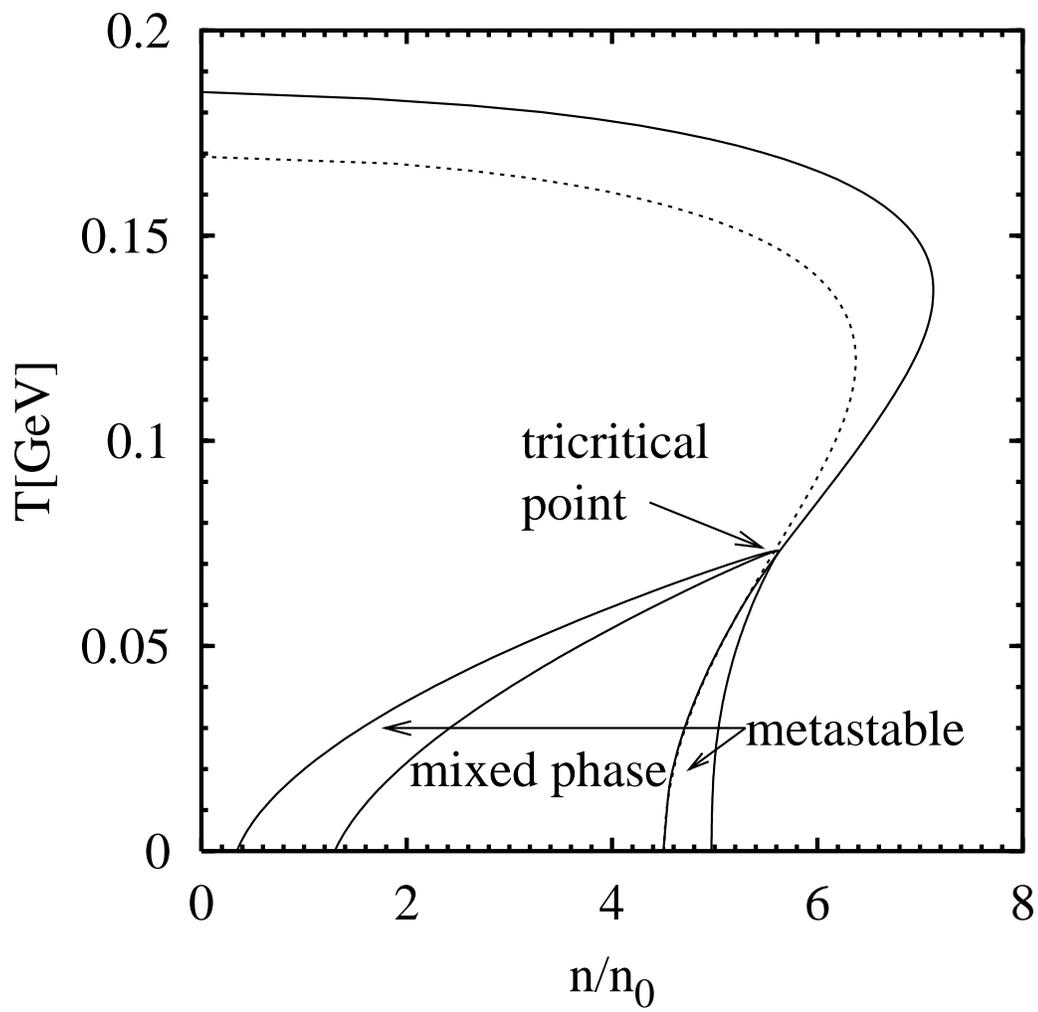
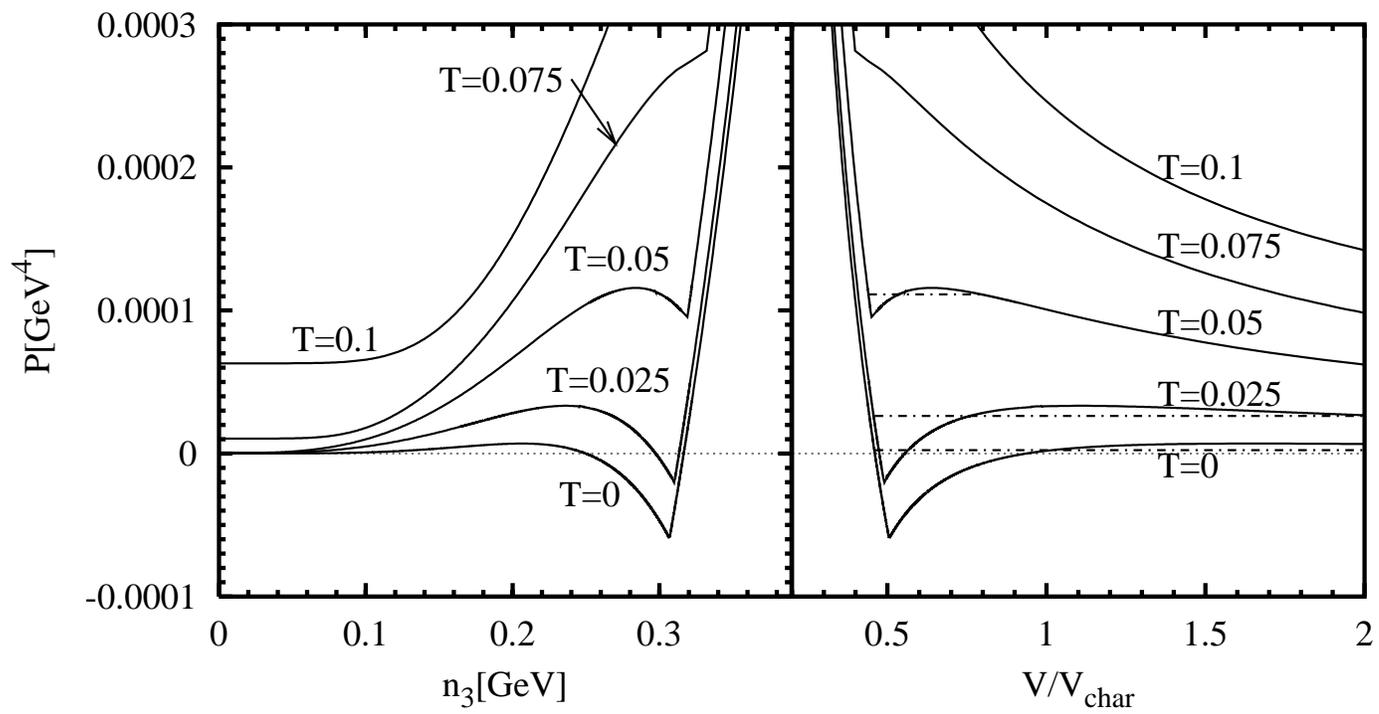
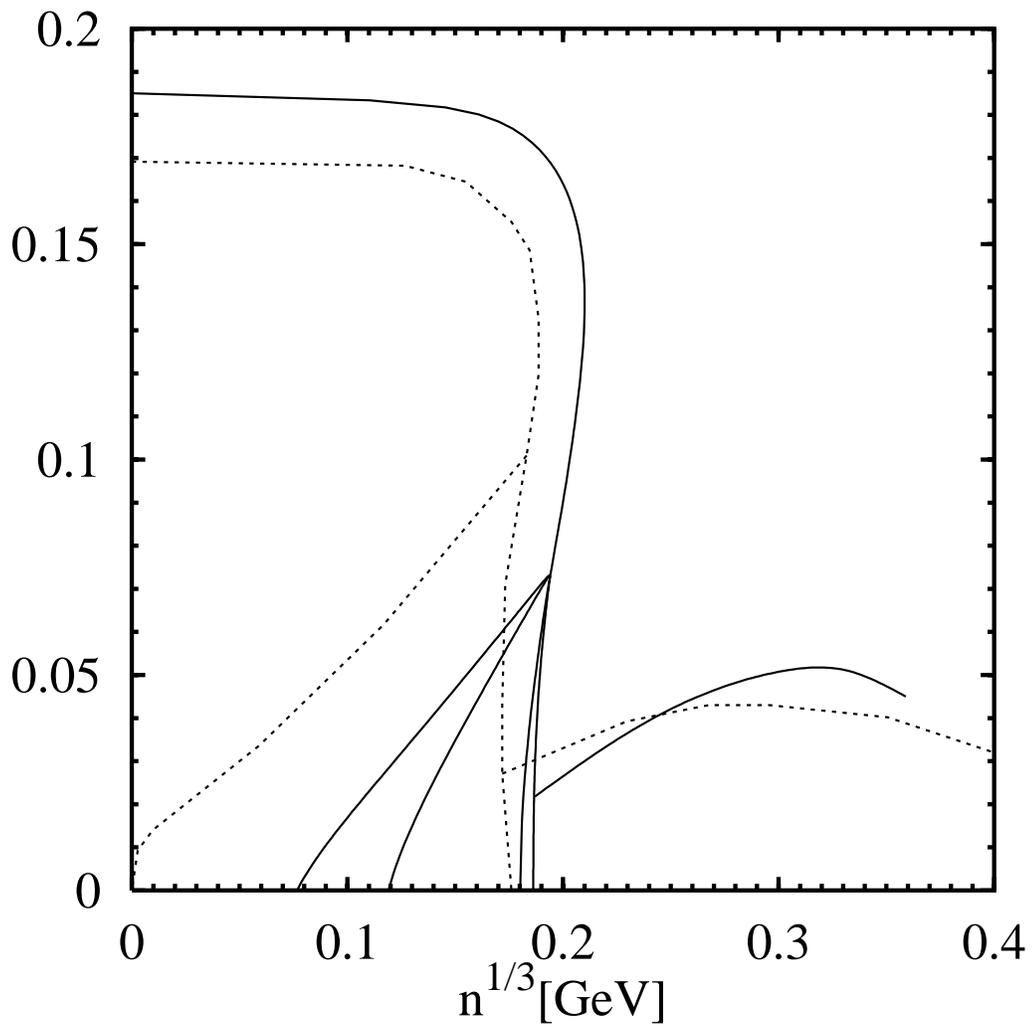
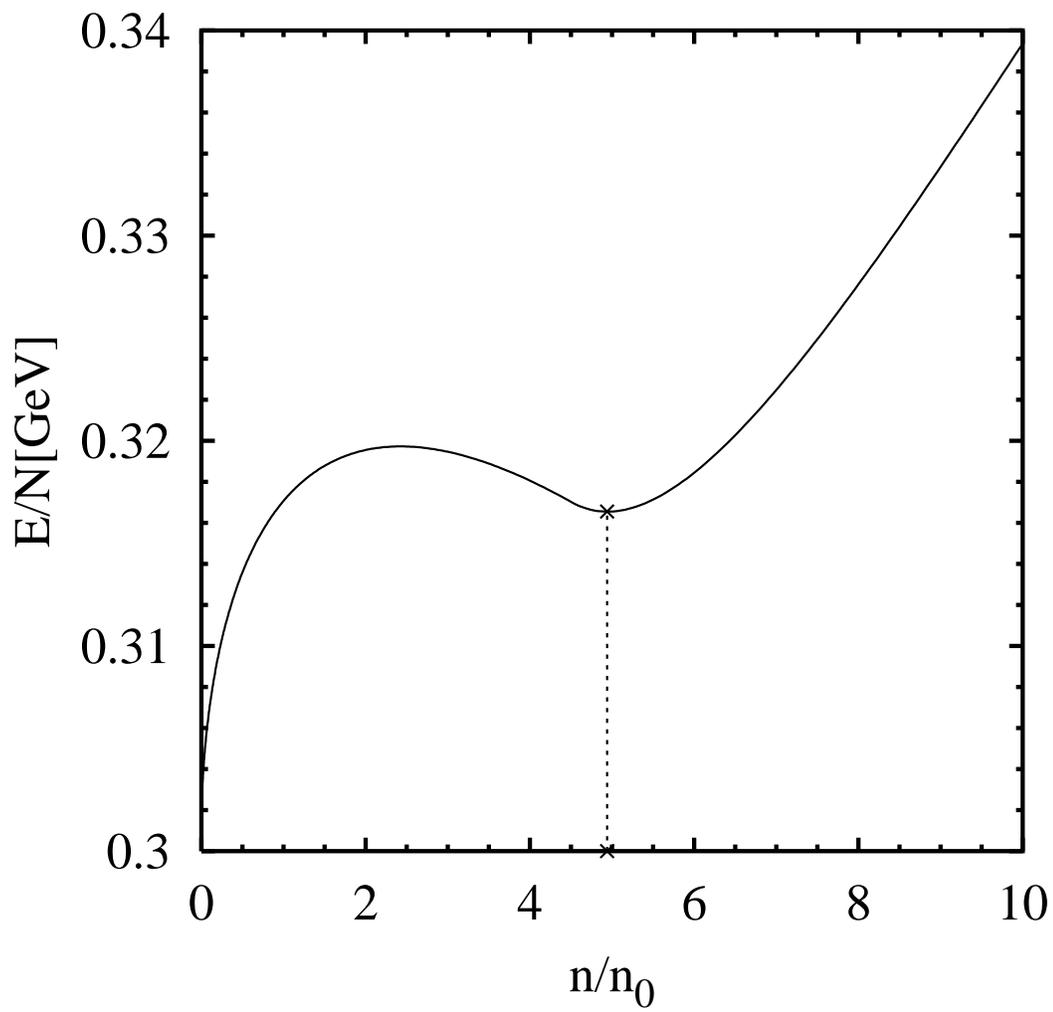
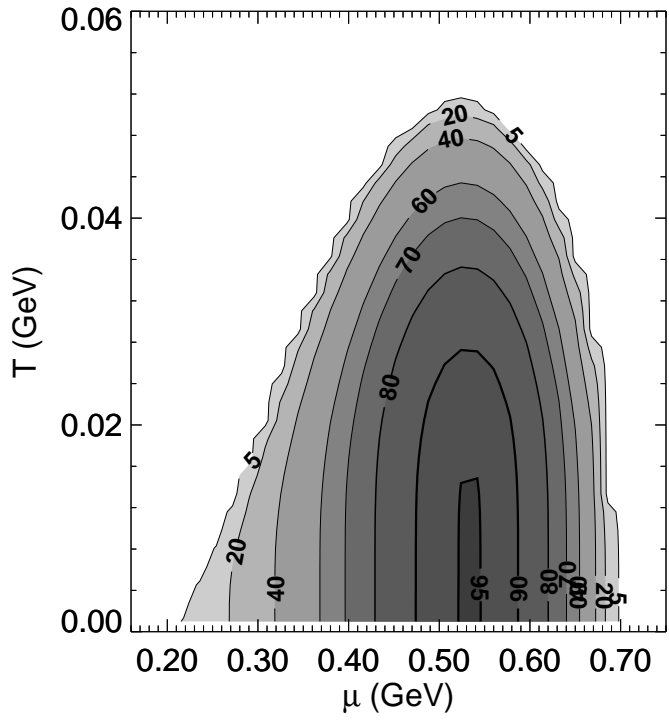
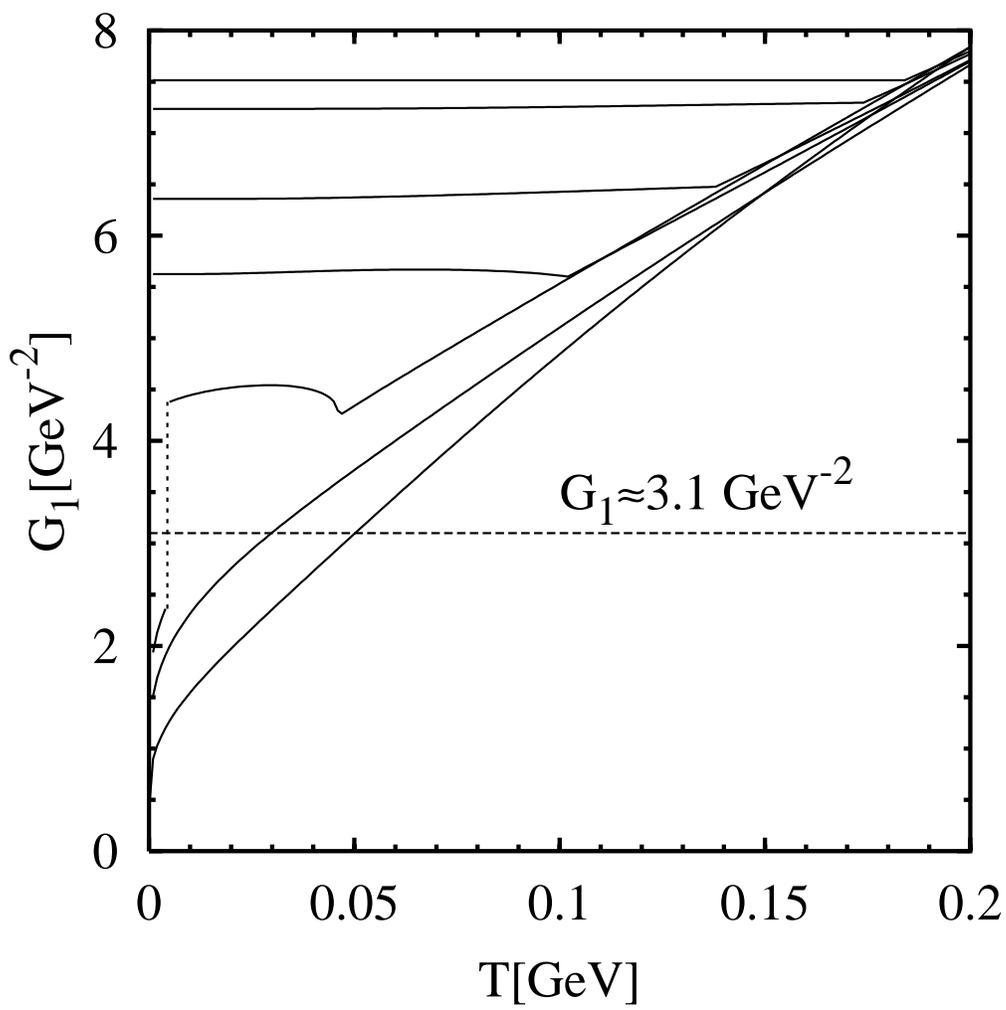
\end{document}